\documentclass[aps,pre,twocolumn,a4paper,footinbib,longbibliography,showpacs,groupedaddress,floatfix,superscriptaddress]{revtex4-2}%

\usepackage[dvipsnames]{xcolor}
\definecolor{linkcolor}{rgb}{0.3,0.3,1.0} 
\usepackage[pdftex,colorlinks=true, linkcolor= linkcolor, citecolor= linkcolor, urlcolor= linkcolor, hyperindex=true,hyperfigures=true]{hyperref} 

\usepackage{amssymb,amsmath,amsfonts}
\usepackage{graphicx}
\usepackage{float}
\usepackage{bm}
\usepackage{physics}
\usepackage{mathtools}

\usepackage{comment}

\begin{document}

\title{Refining Unified Colored-Noise Approximation}
\author{Alessandro Corso}
\affiliation{Department of Physics and Astronomy ``G. Galilei'', University of Padova, 35131 Padova, Italy}
\affiliation{Galilean School of Higher Education, 35131 Padova, Italy}
\author{Daniel M. Busiello}
\affiliation{Department of Physics and Astronomy ``G. Galilei'', University of Padova, 35131 Padova, Italy}
\affiliation{Max Planck Institute for the Physics of Complex Systems, 01187 Dresden, Germany}
\affiliation{Theoretical Sciences Visiting Program (TSVP), Okinawa Institute of Science and Technology Graduate University, Onna, 904-0495, Japan}

\begin{abstract}
    Countless biological and physical systems experience fluctuations that exhibit non-trivial temporal correlations. The Unified Colored-Noise Approximation (UCNA) is a framework providing an approximate description of such stochastic dynamics with colored noise, valid in the limits of vanishing and infinite correlation time of the noise. We first pinpoint and address some criticalities in its original derivation, recasting the result through a time-scale separation procedure. By using our approach, we derive the next-to-leading order correction to the dynamics in both limiting regimes, and highlight the relevant physical scalings of these approximations. Our result helps frame the limits of validity of both the original and the refined formulas, especially in comparison with those derived through different approximation procedures. We show our findings in two paradigmatic examples, a quartic potential and a stochastic logistic growth with multiplicative noise.
\end{abstract}

\maketitle

\section{Introduction}
The presence of white noise is a common assumption in many applications involving stochastic dynamics, and is often justified by thermal fluctuations or environmental noise with negligible correlation times \cite{haunggi1994colored,peliti2021stochastic,gardiner2004handbook,seifert2012stochastic}. Nevertheless, the description of many real-world complex systems requires the presence of noise with finite correlation time, i.e., a colored noise. This is the case, for example, of degrees of freedom in contact with active baths where dissipative processes occur over finite time scales \cite{Caprini_2022_active_particles,Paoluzzi_2016_critical_active_matter,Wittmann_2017_active_matter_I,dabelow2019irreversibility}; beads immersed in a polymer solution, where the complex internal viscoelastic structure of polymers is reflected in frequency-dependent noise \cite{gittes1997microscopic,di2020explicit,venturelli2023memory}; and ecological dynamics subject to changing interactions and correlated environmental fluctuations \cite{morales1999viability,PRL2024_Azaele,Spanio_2017_colored_population_dynamics}. 
In all these examples, the emergence of colored noise is associated with the presence of hidden degrees of freedom evolving on characteristic time scales \cite{busiello2024unraveling,maes2020fluctuating}. Therefore, the white-noise assumption is recovered for a given observable only when its dynamics exhibits an intrinsic time scale much longer than this noise correlation time induced by these hidden variables \cite{ucna_jung_hanggi_1987}.

A direct and comprehensive analytical treatment of stochastic equations in the presence of colored noise is usually out of reach. The derivation of estimates for physically relevant observables, i.e., stationary distributions or escape rates, often resorts to various approximation schemes. For instance, the Best Fokker-Plank Equation (BFPE), valid in the weak-noise limit, has been derived both by means of a functional calculus approach \cite{sancho_sanmiguel_1982} and through a cumulant-perturbation technique \cite{Bianucci_2020}. The Local Linearization Assumption (LLA), which may be formally interpreted as a short-correlation-time expansion of the BFPE \cite{Bianucci_2024_pitfall}, can also be obtained through a different truncation of the exact Kramers-Moyal expansion \cite{Fox_1986_functional}. Several other approaches have been proposed, even beyond the context of colored noise \cite{graham1984weak,graham1985weak,hanggi1990reaction,remlein2022coherence}, with the majority of them tailored to the limit of weak noise or short noise correlation time. The Unified Colored-Noise Approximation (UCNA) stands out in this respect, as its validity extends to both short and long correlation time of the noise, and possibly provides a good description also for intermediate regimes \cite{ucna_jung_hanggi_1987}. Despite the UCNA has recently found numerous applications ranging from active matter \cite{o2022time,maggi2015multidimensional,caprini2022active} to population dynamics \cite{zanchetta2025emergence,Spanio_2017_colored_population_dynamics}, some fundamental problems remain open regarding its range of application, its derivation, and its next-to-leading order corrections. In fact, the original formulation of the UCNA \cite{ucna_jung_hanggi_1987} does not solve some mathematical ambiguities regarding the scaling of the diffusion coefficient, thereby relying on a naive adiabatic elimination of the noise variable. It is thus hard to identify the order of the terms that are being ignored during the procedure, assess its connection to other approximation schemes, and derive higher-order correction terms.

In this work, we pinpoint and address the criticalities associated with the original formulation of the UCNA, proposing an approach based on a time-scale separation in Section ~\ref{sec:timescale_sep}. By employing our framework, we also analytically derive the next-to-leading order correction to the UCNA in both small and large noise correlation time in Section ~\ref{sec:next_order}. The proposed approach allows us to highlight the relevant scaling of the approximation terms and directly compare the UCNA to different approximation procedures. In particular, we focus on the Local Linearization Approach (LLA) since it leads to the same stationary solution as the UCNA at the leading order. Finally, our findings are tested in two paradigmatic examples: a quartic potential (see Section \ref{sec:example_bistable}) and the stochastic logistic growth (see Section \ref{sec:example_logistic}).

\section{UCNA via time-scale separation \label{sec:timescale_sep}}

To retrace the steps of the derivation of the UCNA, we consider a general Langevin equation describing the evolution of a single degree of freedom,
\begin{equation}
    \label{eq: Langevin}
    \dot{x} = f(x) + \eta(t) \;,
\end{equation}
subject to a colored noise $\eta(t)$ with zero mean and correlation time $\tau$, i.e.,
\begin{equation}
    \langle \eta(t) \eta(s) \rangle = \frac{D}{\tau} ~e^{-|t-s|/\tau}\;,
\end{equation}
where $D$ is the diffusion coefficient. In the presence of multiplicative fluctuations, an additive noise can be reconstructed through a suitable change of variables \cite{gardiner2004handbook}. For the sake of simplicity, Equation \eqref{eq: Langevin} is expressed in terms of adimensional variables, i.e., the time has been rescaled by the typical time scale of the system $t \to t ~\tau_s$, while $x\to x L$, where $L$ is a characteristic length-scale. In this way, both $D$ and $\tau$ are dimensionless quantities and can be directly used as expansion parameters. The white noise limit is obtained for $\tau\to 0$.

A possible route to deal with colored noise is to map the system into an equivalent set of two equations driven by white noise:
\begin{eqnarray}
    \label{eq: x color}
    \dot{x} &=& f(x) + \eta \\
    \dot{\eta} &=& - \frac{\eta}{\tau} + \frac{1}{\tau} \xi(t)
    \label{eq: eta color}
\end{eqnarray}
where $\xi(t)$ has zero mean and $\langle \xi(t) \xi(s)\rangle = 2 D \delta(t-s)$. Eq.~\ref{eq: x color} represents the starting point of the adiabatic elimination procedure. In the small-$\tau$ limit, the elimination can be performed both through the formalism of projection operators \cite{gardiner2004handbook} or time-scale separation \cite{busiello2020coarse,bo2017multiple,nicoletti2024information}. Alternative procedures, such as the functional-calculus approach \cite{sancho_sanmiguel_1982} or the cumulant-perturbation technique \cite{Bianucci_2020}, avoid adding an additional variable and instead are based on producing, first, an exact master equation for the system with colored noise, and then truncating it to obtain an approximated Fokker-Plank equation.

Building an approximation that holds in both $\tau\to 0$ and $\tau\to +\infty$ requires the identification of a scaling parameter $\epsilon$ which vanishes in both limits. Moreover, $\epsilon$ has to provide a global scaling, i.e., when expressing the equations in terms of powers of $\epsilon$, no spurious dependences on $\tau$ remain. The UCNA approach prescribes to express $\eta$ using Eq.~\eqref{eq: x color} and substitute it into Eq.~\eqref{eq: eta color} to have a stochastic differential equation for $x$ only:
\begin{equation}
    \tau \ddot{x} = (1 - \tau f'(x)) ~\dot{x} + f(x) + \xi(t)
\end{equation}
with $f'(x) < 0$ in regions of local stability \cite{ucna_jung_hanggi_1987}, with the superscript $\cdot'$ indicating the derivative over $x$. After introducing a rescaled time by $t \rightarrow t \sqrt{\tau} $ one obtains
\begin{equation}
    \ddot{x} = \gamma_\tau(x) ~\dot{x} + f(x) + \xi(t)\;,
\end{equation}
where $\gamma_\tau(x) = (\tau^{-1/2} - \tau^{1/2} f'(x))$ is an effective damping coefficient which diverges for both $\tau\to 0$ and $\tau\to +\infty$, suggesting a naive adiabatic elimination (in the same spirit of the overdamped limit) setting $\ddot{x}\approx 0$. This procedure leads to a Langevin dynamics for $x$ and an associated UCNA Fokker-Planck equation.

The first problem we identify resides in the adiabatic elimination. To make it rigorous, we use a time-scale separation approach starting from Eqs.~\eqref{eq: x color} and \eqref{eq: eta color}. Naming $v = f(x) + \eta$, so that $v$ is the fastest variable to set at stationarity, the Fokker-Planck equation for the joint distribution $p(x,v,t)$ is:
\begin{equation}
    \partial_t p = - v \partial_x p + \partial_v \left( \frac{1-\tau f'(x)}{\tau} v p + \frac{f(x)}{\tau} p \right) + \frac{D}{\tau^2} \partial_{v^2} p \nonumber \;.
\end{equation}
However, starting from this equation, it is easy to convince yourself that there is no change of variables that makes possible to define an $\epsilon$ whose powers rescale all terms, leaving no spurious $\tau$ dependencies. Without taking any further assumption, the UCNA Fokker-Planck equation cannot be rigorously derived.

Let us start again from Eqs.~\eqref{eq: x color} and \eqref{eq: eta color}. We introduce a $\tau$-dependent diffusion coefficient $D_\tau = D\,g(\tau)$ such that $g(\tau)\to 1$ for $\tau\to 0$ and $\tau^{-1} g(\tau) \rightarrow \alpha$ finite as $\tau\rightarrow +\infty $. Although it seems artificial, this addition has a natural physical interpretation. Indeed, while the white noise limit is restored for $\tau\to 0$ as before, we now have
\begin{equation}
    \langle \eta(t) \eta(s) \rangle = \frac{D_\tau}{\tau} e^{-|t-s|/\tau} \xrightarrow{\tau\to +\infty} \alpha D
\end{equation}
for finite intervals $(t-s)$, which consistently describes slow varying fluctuations of finite strength.

Writing down the Fokker-Planck equation for $p(x,v,t)$ with this modified diffusion, and introducing the parameter $\kappa = g(\tau)/\sqrt{\tau}$, we obtain:
\begin{equation*}
    \partial_t p = -v \partial_x p -\frac{f(x)}{\tau} \partial_v p + \frac{\kappa \Gamma}{\sqrt{\tau}} \partial_v (v p) + \frac{\kappa D}{\tau^{3/2}} \partial_{v^2} p \;,
\end{equation*}
with $\Gamma = (1-\tau f'(x))/g(\tau)$. Upon rescaling $v\to \nu/\sqrt{\tau}$ and $t\to \theta \sqrt{\tau} \kappa$ (recalling that we are working with dimensionless quantities), and defining the expansion parameter $\epsilon \equiv \kappa^{-1}$, we finally have:
\begin{equation}
    \label{eq: 2Depsilon FPE}
    \epsilon^2 \partial_\theta p = \epsilon \left( \nu \partial_x p + f(x) \partial_\nu p \right) + \Gamma \,\partial_\nu \left(\left( \nu + \frac{D}{\Gamma} \right) p \right) \,.
\end{equation}
Notice that, in Eq.~\eqref{eq: 2Depsilon FPE}, the scaling with respect to $\tau$ in both limits is solely controlled by $\epsilon$. All residual dependencies on $\tau$ are in $\Gamma$, but this function assumes finite values for all $\tau$.

The time-scale separation can now be performed on Eq.~\eqref{eq: 2Depsilon FPE} (see Appendix \ref{app:derivation} and \cite{nicoletti2025stochastic}), resulting in a solution of the form
\begin{equation}
    p(x,v,t) = \hat{p}_0(x,t) \pi(x,v) + \mathcal{O}(\epsilon)
\end{equation}
with $\int dv \,  \pi(x,v) = 1 $ and $\hat{p}_0$, the real-space marginal up to order $\mathcal{O}(\epsilon)$,  satisfying precisely the UCNA Fokker-Planck equation in the original (non-rescaled) variables:
\begin{equation}
    \label{eq: UCNA FPE}
    \partial_t \hat{p}_0 = -\partial_x \left( \frac{f(x)}{\Gamma_0} \hat{p}_0 \right) + D^{\rm obs} \partial_x \left( \frac{1}{\Gamma_0} \partial_x \left( \frac{\hat{p}_0}{\Gamma_0} \right) \right)
\end{equation}
with $\Gamma_0 \equiv \Gamma\,g(\tau) = 1 - \tau f'(x)$, and $D^{\rm obs} = D_\tau$, compatibly with the proposed change in the starting dynamics. Indeed, according to our definition, at any given $\tau$, $D_\tau$ would be the observed value for the diffusion coefficient.

\section{Next-to-leading correction \label{sec:next_order}}

A main advantage of our procedure is the possibility to proceed further in the $\epsilon$-expansion, thereby deriving the next-to-leading order correction to the UCNA and highlighting the relevant physical scaling of this approximation scheme.

We first propose a general solution of the form:
\begin{equation}
    p(x,v,t) = p_0 + \epsilon\,p_1 + \epsilon^2 p_2 + \mathcal{O}(\epsilon^3) \;.
\end{equation}
Analogously to the zeroth-order solution, we find that $p_1$ can be decomposed as $\hat{p}_1(x,t) \pi(x,v)$, with $\hat{p}_1$ satisfying an equation of the same form of Eq.~\eqref{eq: UCNA FPE}. Therefore, we obtain our first result: the UCNA Fokker-Planck equation remains valid up to $\mathcal{O}(\epsilon^2)$. Thus, the stationary solution up to the first-order of the expansion $\hat{p}_1^{\rm st}$, in the original variables, reads:
\begin{equation}
    \label{eq: UCNA steady}
    \hat{p}_1^{\rm st}(x) = \frac{1}{Z} \,\Gamma_0(x) \,{\rm exp}\left( \frac{1}{D^{\rm obs}} \int^x f(y) \,\Gamma_0(y) \,dy \right)
\end{equation}
where $Z$ is the normalization factor. This finding sheds light on the robustness of the UCNA approximation. In Appendix \ref{app:derivation}, we detail the mathematical derivation of these steps.

To obtain the next relevant correction term to Eq.~\eqref{eq: UCNA FPE}, we proceed further with the expansion to find the equation governing the dynamics in the $x$-space up to $\epsilon^2$. By naming $\mathcal{L}_{\rm UCNA}$ the operator encoding the UCNA Fokker-Planck equation, we arrive at the following form for the \textit{corrected} evolution (see Appendix \ref{app:derivation}):
\begin{equation}
    \label{eq: UCNA corrected}
    \partial_t \hat{p} = \mathcal{L}_{\rm UCNA} \,\hat{p} + \tau \,\mathcal{L}_2(\tau) \,\hat{p}
\end{equation}
where the correction operator $\mathcal{L}_2$ exhibits the following dependence on $\tau$:
\begin{eqnarray}
    \label{eq: scaling L2}
    \mathcal{L}_2(\tau) &\sim& \Gamma^{-3}_0(\tau) \,\mathcal{L}_2^{(1)}(x) + D_\tau \,\Gamma^{-4}_0(\tau) \,\mathcal{L}_2^{(2)}(x) \nonumber \\
    &\;& + \,D_\tau^2 \,\Gamma_0^{-5}(\tau) \,\mathcal{L}_2^{(3)}(x)
\end{eqnarray}
with the expressions for $\mathcal{L}_2^{(1)}(x)$, $\mathcal{L}_2^{(2)}(x)$, and $\mathcal{L}_2^{(3)}(x)$ presented in Appendix \ref{app:derivation}. This is the second main result of this work.

The stationary solution to Eq.~\eqref{eq: UCNA corrected} is usually hard to obtain, hence we resort to a perturbative approach that is valid when the correction is small. As shown in Appendix \ref{app:stationary}, we obtain:
\begin{equation}
    \label{eq: corrected steady}
    \hat{p}^{\rm st}_2(x) \approx \frac{1}{Z_2} \, \hat{p}^{\rm st}_1(x) + \frac{1}{D^{\rm obs}} \int^x dy \, h(y) \, \Gamma_0(y) \,\frac{\hat{p}^{\rm st}_1(x)}{\hat{p}^{\rm st}_1(y)}
\end{equation}
with $Z_2$ ensuring proper normalization.

\begin{figure}[t]
    \centering
    \includegraphics[width = \columnwidth]{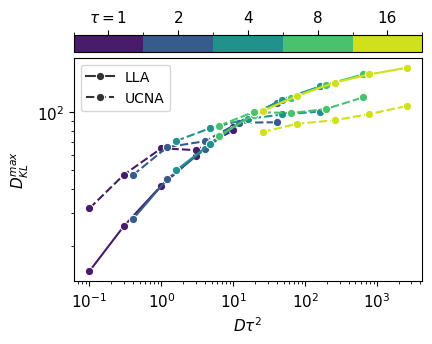}
    \caption{Performance of UCNA and LLA for the quartic potential in Eq.~\eqref{eq: quartic} in the transient regime. The maximum value attained by the Kullback-Leibler divergence between numerical and approximate solutions during the dynamics is used as performance parameter. While the LLA performs better for small $D\tau^2$, we observe a crossover at large values of $D\tau^2$ since the UCNA holds also for $\tau\to +\infty$.}
    \label{fig: transient}
\end{figure}

\subsection{Scaling of UCNA and correction terms}

Eqs.~\eqref{eq: UCNA corrected} and \eqref{eq: scaling L2} highlight the scaling of these corrections with respect to physical parameters. Indeed, notice that $\Gamma_0\to 1$ and $D_\tau\to D$ when $\tau\to 0$; however, for $\tau\to \infty$, $\Gamma_0\sim \tau$ and $D_\tau\sim \tau$. Therefore, by computing the global scaling of $\tau \mathcal{L}_2$, we determine the validity of the UCNA Fokker-Planck equation as a function of $\tau$:
\begin{equation}
    \label{eq: UCNA scaling}
    \partial_t \hat{p} = 
    \begin{cases}
        \mathcal{L}_{\rm UCNA} \,\hat{p} + \mathcal{O}(\tau) \qquad \tau\to 0\\
        \mathcal{L}_{\rm UCNA} \,\hat{p} + \mathcal{O}(\tau^{-2}) \quad \tau\to +\infty
    \end{cases}
\end{equation}
and, for the corrected evolution,
\begin{equation*}
    \partial_t \hat{p} = 
    \begin{cases}
        \mathcal{L}_{\rm UCNA} \,\hat{p} + \tau \,\mathcal{L}_{2} \,\hat{p} + \mathcal{O}(\tau^{3/2}) \qquad \tau\to 0\\
        \mathcal{L}_{\rm UCNA} \,\hat{p} + \tau \,\mathcal{L}_{2} \,\hat{p} + \mathcal{O}(\tau^{-5/2}) \;\,\quad \tau\to +\infty
    \end{cases}
\end{equation*}
As for the dependence on the diffusion coefficient, the zeroth order UCNA operator exhibits a term proportional to $D$, while the correction operator $\mathcal{L}_2$ contains terms proportional to $D^0=1$, $D$, and $D^2$. At the same time, the LLA Fokker-Plank equation, that leads to the same stationary solution as the zeroth order UCNA, Eq.~\eqref{eq: UCNA steady}, is formally valid in the weak-noise and short-correlation-time limits and correct up to $\mathcal{O}(D\tau^2)$. However, previous studies have shown, through numerical simulations, that the LLA may provide very good results even for large noise intensity \cite{Bianucci_2024_pitfall}. An important additional note is that, when evaluating the correction to the UCNA stationary solution using a perturbative approach, Eq.~\eqref{eq: corrected steady}, we introduce an additional source of error on the next-to-leading order term.

\begin{figure*}[t]
    \centering
    \includegraphics[width=0.44\textwidth]{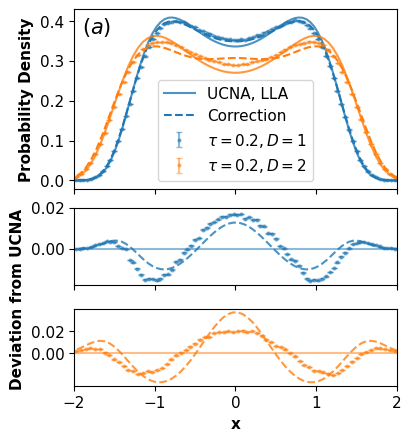}
    \includegraphics[width=0.55\textwidth]{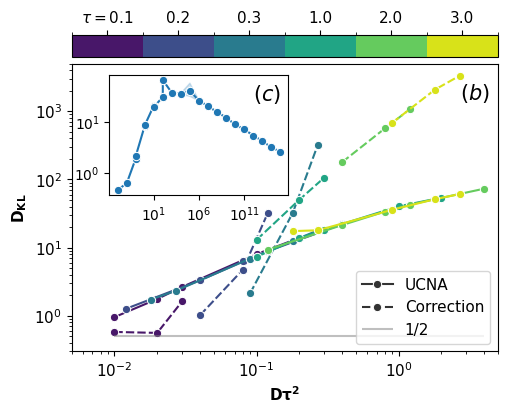}
    \caption{Performance of the theory for the quartic potential in Eq.~\eqref{eq: quartic}. 
    (a) Stationary probability distribution in the UCNA approximation and its correction, compared with simulations of the exact stochastic differential equation. Lower panels show the difference between the two approximations. 
    (b) Kullback-Leibler divergence, $D_{\rm KL}$, in Eq.~\eqref{eq: KL} between numerical solution and approximate theoretical results. The inset (c) shows the behavior of the $D_{\rm KL}$ for large values of $D\tau^2$.}
    \label{fig: quartic}
\end{figure*}

\section{Examples}

In this section, we apply our results to two examples, a particle diffusing in a quartic potential and the stochastic logistic growth. We show how the difference in the range of validity between UCNA, its corrected version, and LLA emerges in the transient regime and, as expected, for large $\tau$.

\subsection{Quartic potential \label{sec:example_bistable}}

The first model we analyze in detail is the motion of a Brownian particle in a quartic potential in the presence of colored noise. This system is described by the following Langevin equation:
\begin{equation}
    \label{eq: quartic}
    \dot{x} = - x^3 + \eta(t) \qquad \langle \eta(t) \eta(s) \rangle = \frac{D}{\tau} e^{-|t-s|/\tau} \;.
\end{equation}

By using our approach, we obtain two approximate forms for the probability distribution of Eq.~\eqref{eq: quartic}, $\hat{p} = \hat{p}_1$ which coincides with the UCNA solution, and $\hat{p}_2$ which incorporates the next-to-leading order correction. To evaluate the performance of our method, we first simulate the dynamics and compute the resulting distribution $p^{\rm sim}$ from the histogram of the counts, dividing the interval into $M$ bins of size $dx$. In particular, naming $N$ the total counts and $N_i$ the counts in each bin, $p_i^{\rm sim} = N_i/N$. Thus, we employ the Kullback-Leibler divergence between $p^{\rm sim}$ and our approximation of order $n$, that is
\begin{equation}
    \label{eq: KL}
    D_{\rm KL}(\hat{p}_n||p^{\rm sim}) = \frac{N}{M} \sum_{i=1}^M \hat{p}_n(x_i) \,dx \,\ln \left( \frac{\hat{p}_n(x_i) \,dx}{p^{\rm sim}_i} \right).
\end{equation}
In the limit of large $N$ and $M$, $D_{KL}(\hat p|| p^{\rm sim})\to 1/2$ if the data in $p^{\rm sim}$ is drawn exactly from $\hat{p}$. We use this as a reference value for our results.

We first compare UCNA and LLA during the transient regime for increasing values of $\tau$ and $D$, using the maximum value attained by $D_{\rm KL}$ during the dynamics, $D_{\rm KL}^{\rm max}$, as performance parameter. While their steady state distributions coincide, in Figure \ref{fig: transient}, we show that LLA performs better for small $D \tau^2$ when $\tau$ is not small. However, when $D \tau^2$ increases, since the UCNA holds also for $\tau\to +\infty$, we observe a crossover in the performance.

\begin{figure*}[t]
    \centering
    \includegraphics[width=0.66\textwidth]{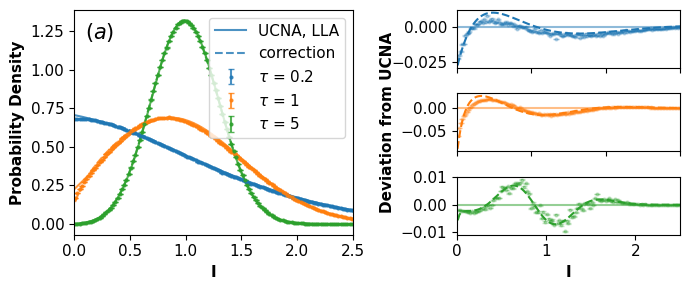}
    \includegraphics[width=0.33\textwidth]{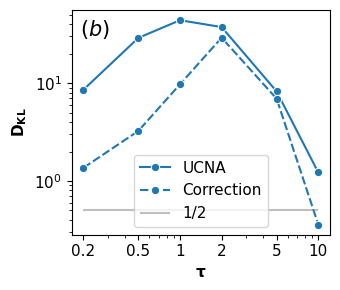}
    \caption{Performance of the theory for the stochastic logistic model at critical noise intensity $D = 2$ for which the steady-state distribution stays finite at $I=0$ (see also \cite{ucna_jung_hanggi_1987}). (a) Stationary probability distribution in the UCNA approximation and its correction, compared with simulations of the exact stochastic differential equation. Side panels show the difference between the two approximations. (b) Kullback-Leibler divergence, $D_{KL}$, between numerical and approximate solution as a function of $\tau$.}
    \label{fig: logistic}
\end{figure*}

Then, we show that the steady-state derived from the corrected UCNA Fokker-Planck equation, Eq.~\eqref{eq: corrected steady}, in the appropriate regime of validity, improves the UCNA, thereby validating our theoretical results. In fact, in Figure \ref{fig: quartic}a, we plot the simulation results (dots) and the analytical expressions (lines). There is a strong agreement when $\tau$ is small enough and $D$ is not too large. As soon as $D$ increases, errors coming from the perturbative solution start to be visible. Figure \ref{fig: quartic}b shows the performance of analytical predictions as a function of $D\tau^2$, since this is the expected scaling in the LLA, which at stationarity coincides with the UCNA. Coherently with our theoretical arguments, the leading contribution to the scaling of the UCNA (and its correction) stems from the increase in $D\tau^2$, while, for very small $D\tau^2$, the additional errors due to the perturbative approach noticeably affect the accuracy of the corrected stationary distribution. The inset confirms the validity of this approach for large $\tau$. The Kullback-Leibler divergence is used as performance parameter.

\subsection{Stochastic logistic growth \label{sec:example_logistic}}

The second example is a logistic growth with environmental noise, a model that has been employed both in population dynamics \cite{descheemaeker2020stochastic,shoemaker2024investigating} and to describe dye laser intensity \cite{ucna_jung_hanggi_1987}. The dynamical equation is:
\begin{equation}
    \dot{I} = 2 \,I \,(1-I)+I\,\eta(t)
\end{equation}
where $\eta(t)$ is the same as in Eq.~\eqref{eq: quartic}. First, we map this system in an equivalent dynamics with additive noise for the variable $x = \log(I)$:
\begin{equation}
    \dot{x} = 2 \,(1-e^x) + \eta(t) \;.
\end{equation}
We see that the next-to-leading order correction to the UCNA improves the accuracy of the stationary solution for three different values of $\tau$ (see Fig.~\ref{fig: logistic}a-b). Importantly, the errors peak at intermediate values, while they reduce for small and large correlation times, as expected (see Fig.~\ref{fig: logistic}c). Also in this case, we employ the Kullback-Leiber divergence in Eq.~\eqref{eq: KL} to quantify the deviations between theoretical and numerical solutions.

\section{Discussion}
In this work, we have highlighted and addressed some criticalities concerning the original formulation of the UCNA, by recasting its derivation in terms of a time-scale separation approach. Through this technique, we shed some light on the range of validity of this approximation, especially in comparison to the LLA. The time-scale separation procedure has also allowed us to derive the next-to-leading order correction to the UCNA. We have demonstrated the accuracy of the proposed correction and its regimes of validity in two paradigmatic examples: a Brownian particle in a quartic potential and a model for logistic growth with environmental noise.

Time-scale separation approaches have recently been used in the context of stochastic thermodynamics, chemical networks, and information theory to gain insights into otherwise very complex dynamics \cite{busiello2020coarse,nicoletti2024information,dabelow2019irreversibility,bo2014entropy,avanzini2023circuit}. However, they rely on expansions that are usually truncated at the leading order, with little to no considerations on the deviations from this leading behavior. Here, by systematically proceeding up to the next-to-leading order in the dynamics, we showed the complexity of the approach and, at the same time, provided a simplified expression to evaluate the resulting correction term by means of a perturbative approach. Therefore, our results might open the way toward a theoretical understanding of a variety of physical and biological systems in which the time-scale separation does not strictly hold, and for which studying only the leading order might be too restrictive. Relevant examples include reaction-diffusion systems in regimes where the Damk{\"o}hler number takes finite values \cite{dass2021equilibrium,brauns2020phase}, ecological models with finite correlation times for the environmental noise \cite{PRL2024_Azaele,azaele2025recent}, and information-theoretic approaches to understanding the emergence of observed behaviors when the time scales of the underlying processes become entangled \cite{fotowat2023neural,nicoletti2025optimal}.

\begin{acknowledgments}
The authors thank Amos Maritan and Sandro Azaele for their insightful discussions and comments throughout the development of this study. Part of this research was conducted during the visit of D.M.B. at the Okinawa Institute of Science and Technology (OIST) through the Theoretical Sciences Visiting Program (TSVP). D.M.B. is funded by the STARS@UNIPD program through the project ``ActiveInfo''.
\end{acknowledgments}

\appendix

\section{Time-scale separation procedure}\label{app:derivation}

We show here in detail how a time-scale separation procedure can be used to obtain the UCNA result and its next-to-leading order correction. Starting from Eq.~\eqref{eq: 2Depsilon FPE},
\begin{align}\label{eq: FP epsilon}
    \epsilon^2 \partial_\theta p = -\epsilon (\nu\partial_x+f\partial_\nu)p+\Gamma\partial_\nu\left(\nu+\frac{D}{\Gamma}\partial_\nu\right)p \;,
\end{align}
we expand the joint probability distribution in the small parameter $\epsilon$, defining $p = p_0 + \epsilon p_1 + \epsilon^2 p_2 + \dots $. The resulting equation can be solved order by order in $\epsilon$. We are interested in the time evolution of the marginalized probability $\hat{p}(x,\theta)=\int p(x,\nu,\theta)d\nu $: since the time derivative in Eq.~\eqref{eq: FP epsilon} appears at $\mathcal{O}(\epsilon^2)$, it is clear that the dynamical term, $\partial_\theta p_0$, will appear at second order, its correction at the third order, and so on.

It is convenient to rewrite the joint probability distribution at each order as  
\begin{align}
    p_i(x,\nu,\theta) &= p_s(x,\nu)\psi_i(x,\nu,\theta) \quad \forall \, i=0,1,\dots \;,
\end{align}
factoring out the kernel of the lowest order operator,
\begin{align*}
    p_s(x,\nu)=\frac{\exp\left(-\frac{\Gamma \nu^2}{2D}\right)}{\sqrt{2\pi D}} \quad\text{s.t.}\quad \partial_\nu\left(\nu+\frac{D}{\Gamma}\partial_\nu\right)p_s=0 \;.
\end{align*}
At zeroth order, Eq.~\eqref{eq: FP epsilon} reads 
\begin{align}
    \partial_\nu (p_s \partial_\nu \psi_0) &=0
\end{align}
which is solved by $\psi_0(x,\nu,\theta) = \phi_0(x,\theta)$. Going further, at first order in $\epsilon$ one obtains
\begin{align}
    D\partial_\nu (p_s \partial_\nu\psi_1) &=  (f\partial_\nu + \nu \partial_x) p_s \phi_0,
\end{align}
so that, integrating twice, 
\begin{align}
    \psi_1 &= \left[\frac{\Gamma'\phi_0}{\Gamma^2}+ \frac{f\phi_0}{D}-\frac{\partial_x\phi_0}{\Gamma}\right] \nu + \frac{\Gamma'\phi_0}{6D\Gamma} \nu^3+ \phi_1
\end{align}
where $\phi_1 $ is constant in $\nu$. Except for this constant, $\psi_1$ is odd in $\nu$, thus only $\phi_1$ survives after marginalization:
\begin{align}  \nonumber 
    \hat{p}(x,\theta) &=  \int_{-\infty}^{+\infty} d\nu \, p_s(x,\nu) \left[\phi_0(x,\theta) + \phi_1(x,\theta)\right] +\mathcal{O}(\epsilon^2) = \\ \label{eq: marginal expansion}
    &= \Gamma^{-1/2}(x) \left[\phi_0(x,\theta) + \phi_1(x,\theta)\right] +\mathcal{O}(\epsilon^2) \;.
\end{align}
At the second order, one obtains an equation for the time evolution of $\phi_0$:
\begin{align}\label{eq: second order}
    D \partial_\nu (p_s \partial_\nu \psi_2) &= (f\partial_\nu + \nu \partial_x) p_1 + \partial_\theta p_0 \;,
\end{align}
and integrating out $\nu$,
\begin{align}
    \frac{1}{\sqrt{\Gamma}}\partial_\theta \phi_0 &=-\partial_x\int_{-\infty}^{+\infty} d\nu \, \frac{D}{\Gamma}p_s \partial_\nu\psi_1 \nonumber \\
    \label{eq: zeroth order dynamics}
    &=-\partial_x\left( \frac{f\phi_0 }{\Gamma^{3/2}}\right)+\partial_x\left(\frac{D}{\Gamma}\partial_x\left(\frac{\phi_0 }{\Gamma^{3/2}}\right)\right) \;.
\end{align}
Thus, one arrives at the following dynamics for $\hat p$:
\begin{align}\label{eq: zeroth order ucna}
     \partial_\theta \hat{p} =-\partial_x\left( \frac{f\hat{p}}{\Gamma}\right)+\partial_x\left(\frac{D}{\Gamma}\partial_x\left(\frac{\hat{p}}{\Gamma}\right)\right) + \mathcal{O}(\epsilon)
\end{align} 
Eq.~\eqref{eq: zeroth order dynamics} can be easily recognized as the UCNA Fokker-Plank equation in Eq.~\eqref{eq: UCNA FPE}.

Our time-scale separation procedure allows us to proceed systematically to higher orders in $\epsilon$. At each order, one can solve the following equation
\begin{align}\label{eq: generic order}
    D \partial_\nu (p_s \partial_\nu \psi_i) &= (f\partial_\nu + \nu \partial_x) p_s\psi_{i-1} + \partial_\theta p_s \psi_{i-2}
\end{align}
to find $\psi_i$, by integrating twice over $\nu$. At the third order, one determines the dynamics of $\phi_1$, which is an equation identical to Eq.~\eqref{eq: zeroth order dynamics}:
\begin{align}\label{eq: first order dynamics}
    \frac{1}{\sqrt{\Gamma}}\partial_\theta \phi_1 = -\partial_x\left( \frac{f\phi_1 }{\Gamma^{3/2}}\right)+\partial_x\left(\frac{D}{\Gamma}\partial_x\left(\frac{\phi_1 }{\Gamma^{3/2}}\right)\right) \,.
\end{align}
Therefore, given the definition in Eq.~\eqref{eq: marginal expansion}, Eqs.~\eqref{eq: zeroth order dynamics} and \eqref{eq: first order dynamics} indicate that the UCNA Fokker-Plank, Eq.~\eqref{eq: UCNA FPE}, is valid up to $\mathcal{O}(\epsilon^2)$.

Going further, one obtains the first correction to the UCNA form. Indeed, the dynamics of the second order correction to the probability distribution is, in fact, 
\begin{equation}
    \partial_\theta \hat{p}_2 = \mathcal{L}_{\rm UCNA} \,\hat{p}_2 + \epsilon^2 \,\mathcal{L}_2 \,\hat{p}_0
\end{equation}
where the correction operator is, explicitly, 
\begin{widetext}
    \begin{align} \nonumber
    \mathcal{L}_2 \hat p_0= \frac{1}{8 \Gamma^{19/2}} 
    &\bigl[   2346 D^2 \Gamma'^4  \phi_0-3 D^2 \Gamma \Gamma'^2 (541 \Gamma''  \phi_0+482 \Gamma'  \partial_x \phi_0)-2 D \Gamma^3 ( \phi_0 (2 D \Gamma^4+ 127 f' \Gamma'^2+ 111 f \Gamma' \Gamma'')+ \\ \nonumber 
    &+ 2 D (6 \Gamma^3  \partial_x \phi_0+11 \Gamma''  \partial_x^2 \phi_0+6 \Gamma'  \partial_x^3\phi_0)+ 116 f \Gamma'^2  \partial_x \phi_0)-4 \Gamma^5 (2 D (f''  \partial_x \phi_0+f'  \partial_x^2 \phi_0)+11 f f' \Gamma'  \phi_0+ \\ \nonumber
    &+2 f^2 (\Gamma''  \phi_0+\Gamma'  \partial_x \phi_0))+4 \Gamma^4 (24 D f' \Gamma'  \partial_x \phi_0+D  \phi_0 (8 f'' \Gamma'+11 f' \Gamma'')+ \\ \nonumber
    &+D f (3 \Gamma^3  \phi_0+10 \Gamma''  \partial_x \phi_0+7 \Gamma'  \partial_x^2 \phi_0)+ 9 f^2 \Gamma'^2  \phi_0)+ D \Gamma^2 (4 D \Gamma' (137 \Gamma''  \partial_x \phi_0+78 \Gamma'  \partial_x^2 \phi_0)+ \\ \label{eq: correction operator explicit}
    &+ 2 D  \phi_0 (53 \Gamma''^2+66 \Gamma^3 \Gamma')+ 507 f \Gamma'^3  \phi_0)+ 8 \Gamma^6 (f f'  \partial_x \phi_0+ \phi_0 (f f''+f'^2))
    \bigr] \;.
\end{align}
\end{widetext}

\section{Corrected stationary distribution}\label{app:stationary}

The corrected stationary equation is
\begin{align}\label{eq: corrected stationary appendix}
    0 =& \frac{f\hat{p}}{\Gamma_0} - \frac{D}{\Gamma_0} \partial_x\left(\frac{\hat{p}}{\Gamma_0}\right) + \tau \mathcal{L}_{2}^{\text{st}} \hat{p}
\end{align}
where $\mathcal{L}_{2}^{\text{st}}$ is the correction operator to the stationary equation, $\mathcal{L}_{2} = -\partial_x \mathcal{L}_{2}^{\text{st}}$, transformed back in the original variables. Since an analytical solution is out of reach, we resort to a perturbative approach. Assuming the correction is small, we can evaluate the correction term on the zeroth order result $\hat{p}_1^{\text{st}}$, for which we have an explicit solution in Eq.~\ref{eq: UCNA steady}: defining for convenience the source 
\begin{align}
    h(x) = \tau \Gamma_0 \, \mathcal{L}_{2}^{\text{st}} \, \hat{p}_1^{\text{st}}(x),
\end{align}
Eq.~\ref{eq: corrected stationary appendix} becomes
\begin{align*}
    0 \approx &f\hat{p} - D \partial_x\left(\frac{\hat{p}}{\Gamma_0}\right) + h.
\end{align*}
Solving the previous equation gives the stationary distribution at the next order in perturbation theory:
\begin{align}
    \hat{p}^{\rm st}_2(x) \approx \frac{1}{Z_2} \, \hat{p}^{\rm st}_1(x) + \frac{1}{D^{\rm obs}} \int^x dy \, h(y) \, \Gamma_0(y) \,\frac{\hat{p}^{\rm st}_1(x)}{\hat{p}^{\rm st}_1(y)}
\end{align}
with $Z_2$ ensuring normalization.

\bibliography{Sources}

\end{document}